\documentclass[12pt,reqno]{amsart}
\usepackage{amsfonts}
\usepackage{amssymb}
\usepackage{amsmath}
\usepackage{amsthm}
\usepackage{setspace}

\newcommand{\re}{{\mathbb R}}    
\newcommand{\cn}{{\mathbb C}}    
\newcommand{\E}{{\mathbb E}}     

\newcommand{\expvBig}[1]{\operatorname{\E}\Bigl[#1\Bigr]}    
\newcommand{\expvbig}[1]{\operatorname{\E}\bigl[#1\bigr]}    

\newtheorem{theorem}{Theorem}

\newtheorem{notation}{Notation}

\theoremstyle{definition}
\newtheorem{definition}[theorem]{Definition}

\theoremstyle{remark}
\newtheorem{remark}[theorem]{Remark}

\begin{document}
\doublespacing
\bibliographystyle{abbrv}

\title{Affine Models}
\author{Christa Cuchiero, Damir Filipovic, Josef Teichmann} 
\address{Vienna Institute of Finance, University of Vienna, and Vienna University of Economics and Business Administration, Heiligenst\"adter Strasse 46-48, A-1190 Wien, Austria; Vienna University of Technology, Department of Mathematical Methods in Economics, Wiedner Hauptstrasse 8--10, A-1040 Wien, Austria}
\email{cuchiero@fam.tuwien.ac.at, damir.filipovic@vif.ac.at,\newline jteichma@fam.tuwien.ac.at}
\thanks{The first and third author gratefully acknowledge the support from the FWF-grant Y 328
(START prize from the Austrian Science Fund). The second author gratefully acknowledges the support from WWTF (Vienna Science and Technology Fund).}

\begin{abstract}
Affine term structure models have gained significant attention in the finance literature, mainly due to their analytical tractability and statistical flexibility. The aim of this article is to present both theoretical foundations as well as empirical aspects of the affine model class. Starting from the original one-factor short-rate models of Vasi\v cek and Cox \emph{et al,} we provide an overview of the properties of regular affine processes and explain their relationship to affine term structure models. Methods for securities pricing and for parameter estimation are also discussed, demonstrating how the analytical tractability of affine models can be exploited for practical purposes.
\end{abstract}
\maketitle
\pagebreak
\section{Definition}

\begin{notation}
Throughout the article, $\langle \cdot,\cdot \rangle$ denotes the standard scalar product on $\re^N$.
\end{notation}

\begin{definition}
Let $r_t$ be a short rate model specified as an affine function of an N-dimensional Markov process $X_t$ with state space $D \subseteq \re^N$:
\begin{align}\label{eq:short-rate}
r_t=l+\langle \lambda, X_t \rangle,
\end{align}
for some (non time-dependent) constants $l \in \re$ and $\lambda \in \re^N$. This is called an \emph{Affine Term Structure Model (ATSM)} if the zero-coupon bond price has exponential affine form, i.e.
\begin{align}\label{eq:affine_bond}
P(t,T)=\expvBig{e^{-\int_t^T r_s ds}\Big|X_t}=e^{G(t,T)+\langle H(t,T),X_t \rangle},
\end{align}
where $\mathbb{E}$ denotes the expectation under a risk neutral probability measure.
\end{definition}

\section{Early Examples}

Early well-known examples are the Vasi\v cek~\cite{vasicek} and the Cox, Ingersoll, Ross~\cite{cir} (see eqf11-024 and eqf11-025) time-homogeneous one-factor short rate models. In~\eqref{eq:short-rate}, both models are characterized by $N=1$, $l=0$ and $\lambda=1$.

\subsection{Vasi\v cek Model:}

$X_t$ follows an Ornstein-Uhlenbeck process on $D=\re$,
\begin {align*}
dX_t=(b+\beta X_t)dt+\sigma dW_t, \quad b, \beta \in \re, \quad \sigma \in \re_+,
\end {align*}
where $W_t$ is a standard Brownian motion. Under these model specifications, bond prices can be explicitly calculated and the corresponding coefficients $G$ and $H$ in~\eqref{eq:affine_bond} are given by
\begin{eqnarray} 
H(t,T)&=&\frac{1-e^{\beta(T-t)}}{\beta},\nonumber\\
G(t,T)&=& \frac{\sigma^2}{2} \int_t^T H^2(s,T) ds + b\int_t^T H(s,T) ds,\nonumber
\end{eqnarray}
provided that $\beta\neq 0$. (see also eqf11-024)
\subsection{Cox-Ingersoll-Ross Model:}

$X_t$ is defined as the solution of the following affine diffusion process on $D=\re_+$, known as \emph{Feller square root process},
\begin{align*}
dX_t=(b+\beta X_t)dt+\sigma\sqrt{X_t}dW_t, \quad b, \sigma \in \re_+, \quad \beta \in \re.
\end{align*}
Like in the Vasi\v cek model, there is a closed-form solution for the bond price. If $\sigma \neq 0$, $G$ and $H$ in ~\eqref{eq:affine_bond} 
are then of the form:
\begin{eqnarray} 
G(t,T)&=&\frac{2 b}{\sigma^2}\ln\Bigg(\frac{2\gamma e^{(\gamma-\beta)(T-t)/2}}{(\gamma-\beta)(e^{\gamma(T-t)}-1)+2\gamma}\Bigg),\nonumber\\
H(t,T)&=&\frac{-2(e^{\gamma(T-t)}-1)}{(\gamma-\beta)(e^{\gamma(T-t)}-1)+2\gamma},\nonumber
\end{eqnarray}
where $\gamma:=\sqrt{\beta^2+2\sigma^2}$. (see also eqf11-025)\\ 
Since the development of these first one-dimensional term-structure models, many multi-factor extensions have been considered with the aim to provide more realistic models.  

\section{Regular Affine Processes}

The generic method to construct ATSMs is to use regular affine processes. A concise mathematical foundation was provided by Duffie, Filipovi\'c and Schachermayer~\cite{dfs}. Henceforth, we fix the state space $D=\re_+^m\times\re^{N-m}$, for some $0 \leq m \leq N$.
\begin{definition}
A Markov process $X$ is called \emph{regular affine} if its characteristic function has exponential-affine dependence on the initial state, i.e. for $t \in \re_+$ and $u \in i\re^N$, there exist $\phi(t,u) \in \cn$ and $\psi(t,u) \in \cn^N$, such that for all $x \in D$ 
\begin{align}\label{eq:def_affine}
\expvbig{e^{\langle u, X_t\rangle}\big|X_0=x}=e^{\phi(t,u)+\langle\psi(t,u),x\rangle}.
\end{align}
Moreover, the functions $\phi$ and $\psi$ are continuous in $t$ and $\partial_t^+ \phi(t,u)|_{t=0}$ and $\partial_t^+ \psi(t,u)|_{t=0}$ exist and are continuous at $u=0$.
\end{definition}
Regular affine processes have been defined and completely characterized in~\cite{dfs}. The main result is stated below.

\begin{theorem}\label{th: main theorem}
A regular affine process is a Feller semimartingale with infinitesimal generator 
\begin{eqnarray} 
\mathcal{A}f(x)&=&\sum_{k,l=1}^{N}A_{kl}(x)\frac{\partial^2 f(x)}{\partial x_k \partial x_l}+\langle B(x),\nabla f(x)\rangle-C(x)f(x)\nonumber\\
&+&\int_{D\backslash\{0\}}(f(x+\xi)-f(x)-\langle \nabla f(x),\chi(\xi)\rangle)M(x,d\xi),\label{eq: generator}
\end{eqnarray}
for $f$ in the set of smooth test functions, with
\begin{eqnarray}
A(x)&=&a+\sum_{i=1}^{m}x_i\alpha_i, \quad a, \alpha_i \in \re^{N \times N},
\label{eq: diffusion matrix}\\
B(x)&=&b+\sum_{i=1}^{N}x_i\beta_i, \quad b,\beta_i \in \re^N,
\label{eq: drift}\\
C(x)&=&c+\sum_{i=1}^{m}x_i\gamma_i, \quad c, \gamma_i \in \re_+,\\
M(x,d\xi)&=&m(d\xi)+\sum_{i=1}^{m}x_i\mu_i(d\xi),\label{eq: jump}
\end{eqnarray}
where $m, \mu_i$ are Borel measures on $D\backslash \{0\}$ and $\chi: \re^N \rightarrow \re^N$ some bounded continuous truncation function with $\chi(\xi)=\xi$ in a neighborhood of $0$.
Furthermore, $\phi$ and $\psi$ in~\eqref{eq:def_affine} solve the generalized Riccati equations,
\begin{eqnarray}
\partial_t \phi(t,u)&=&F(\psi(t,u)), \quad \phi(0,u)=0,\label{eq: Riccati1}\\
\partial_t \psi(t,u)&=&R(\psi(t,u)), \quad \psi(0,u)=u,\label{eq: Riccati2}
\end{eqnarray}
with
\begin{eqnarray}
F(u)&=& \langle au,u\rangle+\langle b,u\rangle-c+\int_{D\backslash \{0\}}\Big(e^{\langle u,\xi\rangle}-1-\langle u,\chi(\xi)\rangle\Big)m(d\xi),\nonumber\\
R_i(u)&=&\langle \alpha_i u,u\rangle + \langle \beta_i, u\rangle-\gamma_i+\int_{D\backslash \{0\}}\Big(e^{\langle u,\xi\rangle}-1-\langle u,\chi(\xi)\rangle\Big)\mu_i(d\xi),\nonumber\\
&&\textrm{for } i\in\{1,\ldots,m\},\nonumber\\
R_i(u)&=&\langle \beta_i, u\rangle, \quad \textrm{for } i\in\{m+1,\ldots,N\}. \nonumber
\end{eqnarray}
Conversely, for any choice of admissible parameters $a$, $\alpha_i$, $b$, $\beta_i$, $c$, $\gamma_i$, $m$, $\mu_i$, there exists a unique regular affine process with generator~\eqref{eq: generator}.
\end{theorem}

\begin{remark}
It is worth noting that the infinitesimal generator of every Feller process on $\re^N$ has the form of the above integro-differential operator~\eqref{eq: generator} with some functions $A$, $B$, $C$ and a kernel $M$. The specific characteristic of regular affine processes is that these functions are all affine, as described in~\eqref{eq: diffusion matrix} - \eqref{eq: jump}.\\
Observe furthermore that by the definition of the infinitesimal generator and the form of $F$ and $R$, we have
\begin{eqnarray}
\frac{d}{dt}\expvbig{e^{\langle u, X_t\rangle}\big|X_0=x}\Big|_{t=0_+}&=&\left(\partial_t^+ \phi(t,u)|_{t=0}+\partial_t^+ \psi(t,u)|_{t=0}\right)e^{\langle u,x \rangle}\nonumber\\
&=&\left(F(u)+\langle R(u),x\rangle\right)e^{\langle u,x \rangle}=\mathcal{A}e^{\langle u,x \rangle}.\nonumber
\end{eqnarray}
This gives the link between the form of the operator $\mathcal{A}$ and the functions $F$ and $R$ in the Riccati equations~\eqref{eq: Riccati1} and~\eqref{eq: Riccati2}.
\end{remark}

\begin{remark}
The above parameters satisfy certain admissibility conditions guaranteeing the existence of the process in $D$. These parameter restrictions can be found in Definition 2.6 and equations (2.23)-(2.24) in~\cite{dfs}. We note that admissibility in particular means $\alpha_{i,kl}=0$ for $i,k,l\leq m$ unless $k=l=i$.\\
\end{remark}

\section{Systematic analysis}

\subsection{Regular affine processes and ATSMs}

Regular affine processes generically induce ATSMs. This relation is explicitly stated in the subsequent argument. Under some technical conditions which are specified  in~\cite{dfs} chapter 11, we have for $r_t$ as defined in~\eqref{eq:short-rate},
\begin{align}\label{eq:extended affine}
\expvBig{e^{-\int_0^t r_s ds}e^{\langle u,X_t\rangle}\Big|X_0=x}=e^{\widetilde\phi(t,u)+\langle\widetilde\psi(t,u),x\rangle}\,,
\end{align}
where 
\begin{eqnarray}
\partial_t\widetilde\phi(t,u)&=&\widetilde F(\widetilde\psi(t,u)),\nonumber\\
\partial_t\widetilde\psi(t,u)&=&\widetilde R(\widetilde\psi(t,u)),\nonumber
\end{eqnarray}
with $\widetilde F(u)=F(u)-l$ and $\widetilde R(u)=R(u)-\lambda$. Setting $u=0$ in~\eqref{eq:extended affine}, one immediately gets~\eqref{eq:affine_bond} with $G(t,T)=\widetilde\phi(T-t,0)$ and $H(t,T)=\widetilde\psi(T-t,0)$.

\subsection{Diffusion case}

Conversely, for a class of diffusions 
\begin{align}\label{eq: diffusion}
dX_t=B(X_t)dt+\sigma(X_t)dW_t
\end{align}
on $D$, Duffie and Kan~\cite{duffie} analyzed when~\eqref{eq:affine_bond} implies an affine diffusion matrix $A=\frac{\sigma \sigma^T}{2}$ and an affine drift $B$ of form~\eqref{eq: diffusion matrix} and~\eqref{eq: drift} respectively.

\subsection{One dimensional nonnegative Markov process}

For $D=\re_+$, Filipovi\'c~\cite{filipo} showed that ~\eqref{eq:short-rate} defines an ATSM if and only if $X_t$ is a regular affine process.

\subsection{Relation to Heath-Jarrow-Morton framework}

Filipovi\'c and Teichmann~\cite{filipoteich} established a relation between the Heath-Jarrow-Morton (HJM) framework (see eqf11-022) and ATSMs: Essentially, all generic finite dimensional realizations\footnote{For a precise definition see~\cite{filipoteich}.} 
of a HJM term structure model are time-inhomogeneous ATSMs.

\section{Canonical Representation}

An ATSM stemming from a regular affine diffusion process $X$ on $\re_+^m \times \re^{N-m}$ can be represented in different ways by applying nonsingular affine transformations to $X$. Indeed, for every nonsingular $N\times N$-matrix $K$ and $\kappa \in \re^N$, the transformation $KX+\kappa$ modifies the particular form of~\eqref{eq: diffusion} and the short rate process~\eqref{eq:short-rate}, while observable quantities (e.g. the term structure or bond prices) remain unchanged. In order to group those $N$-dimensional ATSMs generating identical term structures, Dai and Singleton~\cite{daisingleton} found $N+1$ subfamilies $\mathbb{A}_m(N)$, where $0\leq m \leq N$ is the number of state variables actually appearing in the diffusion matrix (i.e. the dimension of the positive half space). For each class, they specified a canonical representation whose diffusion matrix $\sigma \sigma^T$ is of diagonal form with 
\begin{eqnarray}
(\sigma \sigma^T (x))_{kk}=&\left\{ \begin{array}{ll} 
x_k, & k\leq m\\
1+\sum_{i=1}^{m}\lambda_{k,i}x_i & k > m\nonumber
\end{array} \right.,
\end{eqnarray}
where $\lambda_{k,i} \in \re$. For $N\leq 3$ the Dai-Singleton specification comprises all ATSMs generated by regular affine diffusions on $\re^m_+ \times \re^{N-m}$. The general situation $N > 3$ was analyzed by Cheridito, Filipovi\'c and Kimmel~\cite{cfk}. 

\section{Empirical Aspects}

\subsection{Pricing:}

The price of a claim with payoff function $ f(X_t) $ is given by the risk neutral expectation formula
\begin{align*}
\pi(t,x)=\expvBig{e^{-\int_0^t r_s ds}f(X_t)\Big|X_0=x}.
\end{align*}
Suppose $ f $ can be expressed by 
\begin{align}\label{eq: fourier inv}
f(x)=\int_{\re^N} e^{\langle C+i\lambda, x\rangle} \widetilde f(\lambda)d\lambda, \quad \lambda \in \re^N,
\end{align}
for some integrable function $\widetilde f$ and some constant $C\in \re^N$. If, moreover
\begin{align*}
\expvBig{e^{-\int_0^t r_s ds}e^{\langle C,X_t\rangle}\Big|X_0=x}<\infty,
\end{align*}
then~\eqref{eq:extended affine} implies
\begin{eqnarray}
\pi(t,x)&=&\E\bigg[e^{-\int_0^t r_s ds}\bigg(\int_{\re^N}e^{\langle C+ i \lambda, X_t\rangle} \widetilde f(\lambda)d\lambda\bigg)\bigg|X_0=x\bigg]\nonumber\\
&=&\int_{\re^N}\expvBig{e^{-\int_0^t r_s ds}e^{\langle C+i\lambda, X_t \rangle}\Big|X_0=x}\widetilde f(\lambda)d\lambda\nonumber\\
&=&\int_{\re^N}e^{\widetilde\phi(t,C+i\lambda)+\langle\widetilde\psi(t,C+i\lambda),x\rangle}\widetilde f(\lambda)d\lambda.\nonumber
\end{eqnarray}
Hence, the price $\pi(t,x)$ can be computed via numerical integration, since the integrands are in principle known. For instance, in the case $N=1$, the payoff function of a European call $(e^x-e^k)^+$, where $x$ corresponds to the log price of the underlying and $k$ to the log strike price, satisfies~\eqref{eq: fourier inv}. In particular, we have the following integral representation (see~\cite{hubkalkra})
\begin{align*}
(e^x-e^k)^+=\frac{1}{2\pi }\int_{\re}e^{(C+i\lambda)x} \frac{e^{k(1-C-i\lambda)}}{(C+i\lambda)(C+i\lambda-1)}d\lambda.
\end{align*} 
Therefore, the previous formula to compute the price of the call $\pi(t,x)$ is applicable. An alternative approach leading to the same result can be found in Carr and Madan~\cite{carr}.

\subsection{Estimation:} 

Statistical methods to estimate the parameters of ATSMs have been based on maximum likelihood and generalized method of moments.

Concerning maximum likelihood techniques, the conditional log densities entering into the log likelihood function can in general be obtained by inverse Fourier transformation. Since this procedure is computationally costly, several approximations and limited-information estimation strategies have been considered (e.g.~\cite{singleton}). Another possibility is to use closed form expansions of the log likelihood function which are available for general diffusions~\cite{as} and which have been applied to ATSMs. In the case of Gaussian and Cox-Ingersoll-Ross models, one can forgo such techniques, since the log densities are known in closed form (e.g~\cite{pearson}). 

As conditional moments of the form $\E[X_t^m X_{t-s}^n]$ for $m, n \geq 0$ can be computed from the derivatives of the conditional characteristic function and are in general explicitly known up to the solution of the Riccati ODEs~\eqref{eq: Riccati1} and~\eqref{eq: Riccati2}, the generalized method of moments is an alternative to maximum likelihood estimation (e.g.~\cite{andersen}).

\section{Related articles}
\begin{itemize}
\item eqf08-018
\item eqf11-022
\item eqf11-024
\item eqf11-025
\item eqf11-027
\item eqf13-009
\end{itemize}

\end{document}